# Is VIX still the investor fear gauge? Evidence for the US and BRIC markets

Marco Neffelli[1], Marina Resta[1]

[1]Department of Economics and Business Studies, University of Genova, Italy

**Abstract.** We investigate the relationships of the VIX with US and BRIC markets. In detail, we pick up the analysis from the point left off by (Sarwar, 2012), and we focus on the period: Jan 2007 - Feb 2018, thus capturing the relations before, during and after the 2008 financial crisis. Results pinpoint frequent structural breaks in the VIX and suggest an enhancement around 2008 of the fear transmission in response to negative market moves; largely depending on overlaps in trading hours, this has become even stronger post-crisis for the US, while for BRIC countries has gone back towards pre-crisis levels.

## 1 Introduction

Since its inception, the CBOE Volatility Index (VIX) represents a quick and important measure of market sentiment (Whaley, 2000) as it gives an immediate snapshot of market expectations of near-term (next 30 calendar days) volatility conveyed by the S&P500 stock index option prices. The appellation of "fear index", is due to how the

* We are grateful to Gabriele Gaggero for his initial assistance to this work.

VIX reacts to market fluctuations, as the VIX works better in capturing market downtrends. This is because the weighted blend of Call and Put options forming it is mostly used to hedge against market drawdowns (Whaley, 2009), so that the VIX mirrors the investors' demand for hedging and highlights the rise in this demand.

The role of VIX as natural barometer for the riskiness of financial markets has been widely assessed and described in the literature, mainly with a focus on the relationships with the US equity market. In the seminal work of (Fleming et al., 1995), a multivariate regression model investigates the intertemporal relationships between VIX and S&P100 index[1] at various lags and leads. Major findings include: (*i*) the role of VIX as a fear gauge since it exhibits a statistically significant inverse relationship with the US equity index; (*ii*) the asymmetry of VIX in representing the impact of positive and negative US equity index returns. Besides, (Whaley, 2000) describes the behavior of VIX during US equity market turmoil, identifying a kind of alert threshold (30%) that separates high from low volatility periods. Nevertheless, (Whaley, 2009) dampen the role of this threshold, and highlights how the capability of VIX in reflecting the current state of the economy should be accompanied by a deepest analysis of past conditions affecting the market. (Sarwar, 2012a) extends the analysis in (Fleming et al., 1995) investigating the relationships among the VIX and three US indexes, namely the S&P100, the S&P500 and the S&P600, during the period 1992-2011. Results are aligned with previous studies, and underline a negative simultaneous relationship, which tend to increase during more volatile periods and to decrease elsewhere. The role of VIX as a fear gauge for the S&P500 index has been also discussed in (Chiang, 2012) who discovers an asymmetric response of the VIX to negative returns using a bivariate GARCH model with TAR (Threshold Auto-Regression) in the period 2001-2011.

---

[1] The choice of S&P100 index was intentional, since at that time the VIX was based upon this index.

Cross-country and spillover effects between VIX and foreign equity markets have been also discussed in various works. (Sarwar, 2012b) applies the (Fleming et al., 1995) model for testing VIX interactions with BRIC countries in the period 1993-2007. Results underline that VIX is a fear gauge for China and Brazil during the whole period, and for India during the sub-period 1993-1997. A similar investigation is carried on for six European countries in the 1998-2013 timeframe (Sarwar, 2014), confirming the role of VIX as a cross-market fear measure. Furthermore, (Sarwar and Khan, 2017) find evidence of strong connections among VIX spikes and abrupt drawdowns in a bunch of Latin America indexes in the period 2003-2014, thus assessing the role of fear gauge of VIX also for Latin America countries.

From previous rows, however, an open question pertains the role played by the VIX for US and BRIC markets before, during and after a financial crisis. Focusing in the period: Jan 2007 – Feb 2018, this work is aimed to fill in this gap, because not only it includes the 2008 financial crisis, but also it embraces one of the highest high (80.86, 20 Nov 2008) and one of the lowest low (9.14, 3 Nov 2017) ever observed in the history of the index.

Our goals and results go towards at least two directions. First, we aim at shedding some lights on the statistical representation of VIX with a focus on changes in regimes. In particular, we provide evidence of four structural breaks in the VIX mean level: this allows us to distinct the VIX impact before, throughout, after the crisis, and in the six years after the post-crisis period (2012-2018). Notably, within this latter period, the VIX reverted in mean at pre-crisis levels between 2012 and 2016, to stabilize at lowest levels ever in the period 2016-2018. The results highlight a change in regime almost every two years; a big difference in comparison with the period 1993-2007 that encompasses only two breaks (Sarwar, 2012b). This behavior poses several questions about the VIX as well as the relationships with international stock markets. Second, we are interested at checking if and how the fear gauge role of VIX has changed over the past decade with respect to the US and the BRIC financial markets.

The paper is structured as follows. Section 2 is organized into three units, devoted to: (*i*) characterize the data sample; (*ii*) analyze the VIX in search for (eventual) structural breaks; (*iii*) give summary statistics; Section 3 describes basic features of the multivariate regression model employed in the study; Section 4 presents the main results and Section 5 concludes.

## 2 Data and methods

### 2.1 Dataset and Trading Time-zones

We consider the VIX during the period 03/01/2007 – 01/02/2018, for overall number of 2791 daily observations. VIX interactions are studied with respect to the following indexes: S&P500 (SPX - USA); IBOVESPA (IBOV - Brazil); MOEX (IMOEX - Russia); S&P BSE Sensex (BSESN - India); Shanghai c (SHSEC - China). The S&P500 index is a capitalization-weighted index of 500 stocks representing all major US industries. The Brazilian IBOVESPA is a gross total return index of all the most liquid stocks traded on the San Paulo Stock Exchange. The S&P BSE Sensex index is a cap-weighted index representing at best the Indian industries, with components selected to take into account liquidity and industry representation. The Russian MOEX (formerly MICEX) index is a cap-weighted index representing the 50 most liquid Russian stocks traded on the Moscow exchange. Finally, the Shanghai SHSEC is a capitalization-weighted index. The index constituents list comprises all A-shares and B-shares listed on the Shanghai exchange.

The VIX and all the other indexes are retrieved from S&P CapitalIQ database[2]. We use the same indexes as in (Sarwar, 2012b), with the exception of Russian IMOEX that replaces the AK&M index. The rationale of this substitution resides in considering the MOEX more representative of the Russian financial situation than the AK&M which is released by a rating agency. We selected only days in which the VIX has been traded, due to its central role in this analysis. Moreover, we deal with missing observations by filling gaps with linear interpolation of adjacent available observations.

The peculiar feature of this dataset is that opening and closing hours for trading vary for each index: with respect to the UTC time zone, China (SHSEC) opens first and closes before the VIX opening time. The same applies for India (BSESN), while Russia (IMOEX) and Brazil (IBOV) are the only countries with partially (Russia) or nearly (Brazil) overlapping trading windows to the VIX. The VIX and the selected US market (SPX) share the same opening and closing times.

This opening and closing hours setup directly affects the interaction among the VIX and BRIC stock indexes: (Sarwar, 2012b), for example, found strong evidence of relationships among the VIX and the current value of US, China, India and Brazil indexes, indicating that the relation is always stronger within 24 hours. An index like the Chinese SHSEC, that opens before the VIX, however, should not be immediately conditioned by this latter, because investors will react to today's VIX information during next market opening. Therefore, an eventual dependence between VIX and SHSEC should be captured better by lead parameters, because the SHSEC is *ex post* influenced by the fear index. Indeed, we do not expect the VIX to be influenced by emerging markets as BRIC: in this case, however, the relationship should be captured by lag values, since the BRIC indexes release information before the VIX opening hour.

---

[2] https://www.capitaliq.com/

**2.2 A preliminary analysis of VIX in search for structural breaks**

In general, model estimation is a task that can be affected by heteroscedasticity, autocorrelation and multicollinearity in and between the variables under consideration. However, the estimation results may be also influenced by not identified structural breaks: the presence of structural breaks, in fact, indicates changes in the data generating process, so that relationships with other series may consistently vary.

Already (Guo & Wohart, 2006) provide strong statistical evidence that the VIX varies over time with infrequent but significant changes in the mean level: focusing on the timeframe Jan 1990-Dec 2003, this study highlighted three regimes: pre-1992, 1992-1997, and post-1997. Following this track, we run the Bai and Perron test to detect structural breaks in the VIX mean level in the period Jan 2007-February 2018. The test has been managed in accordance to the remarks found in (Bai and Perron, 2003). In detail, we set the trimming parameter to 0.20 to allow for serial correlation and heteroscedasticity in the VIX time series: this means to set the number of maximum breaks to five.

The results were then evaluated according to the rationale explained in next rows. First, in Panel A, we verified with the Double Maximum (UDmax) test statistics if it was possible to reject the assumption of no breaks; if not, we sequentially tested with the F-test statistic the presence of breaks up to the maximum number of five. Second, to verify the adequacy of the findings from Panel A, and to select the proper number of breaks we built various regression models each considering a number of breaks from zero to five, and we reported the result for the best the regression models according to the LWZ (Liu et al., 1997) information criterion. Moreover, in line with (Guo and Wohar, 2006) we applied the test on the VIX raw as well as on the standardized VIX time series. In this latter case, we excluded outliers, i.e. the observations whose absolute value exceeds the mean plus three times the standard deviation. Results are summarized in Table 1.

**Table 1**: Bai and Perron test results for VIX and standardized VIX (Std-VIX) time series.

| Panel A: Bai and Perron Statistics for Tests of Multiple Structural Breaks | | | | | | |
|---|---|---|---|---|---|---|
|  | UDmax[a] | F(1\|0) | F(2\|1) | F(3\|2) | F(4\|3) | F(5\|4) |
| VIX | 20.2190*** | 13.7178*** | 13.011** | 13.3296** | 11.6154* | - |
| Std-VIX | 30.2438*** | 18.0632*** | 17.89*** | 18.0081*** | 14.5189** | - |
| Panel B: Bai and Perron Statistics for the model with number of breaks selected with the LWZ criterion | | | | | | |
|  | LWZ | C1 | C2 | C3 | C4 | C5 |
| VIX | 3.9606 | 19.8356 | 33.3202 | 23.2531 | 15.8426 | 12.4655 |
|  |  | (0.2586)*** | (0.6694)*** | (0.3290)*** | (0.1075)*** | (0.1558)*** |
| Std-VIX | 3.4331 | 19.7867 | 28.7888 | 22.2245 | 14.4917 | 16.9668 |
|  |  | (0.2592)*** | (0.2633)*** | (0.2411)*** | (0.2245)*** | (0.2542)*** |
| Panel C: Bai and Perron Regime Means, End Dates, and 90% and 95% Confidence Intervals | | | | | | |
|  | Regime 1 | Regime 2 | Regime 3 | Regime 4 | Regime 5 | |
| VIX |  |  |  |  |  | |
| End Date | 12/09/2008 | 14/06/2010 | 15/02/2012 | 24/05/2016 | 09/02/2018 | |
| 95% | [03/01/2007; 09/10/2008] | [08/02/2010; 17/09/2015] | [08/12/2011; 13/06/2013] | [11/06/2015; 21/09/2017] |  | |
| 90% | [03/01/2007; 23/09/2008] | [14/04/2010; 27/03/2014] | [06/01/2012; 30/01/2013] | [28/09/2015; 02/05/2017] |  | |
| Std-VIX |  |  |  |  |  | |
| End Date | 08/09/2008 | 22/07/2010 | 07/03/2012 | 24/06/2016 | 09/02/2018 | |
| 95% | [03/01/2007; 18/03/2009] | [29/05/2009; 03/02/2015] | [19/07/2010; 28/11/2014] | [14/03/2016; 14/08/2017] |  | |
| 90% | [03/01/2007; 26/01/2009] | [28/10/2009; 18/10/2013] | [20/09/2011; 23/10/2014] | [20/04/2016; 19/04/2017] |  | |

Table 1 must be read as follows: Panel A contains the Bai and Perron statistics testing for multiple structural breaks. From left to right, we listed: the UDmax test statistic (with null hypothesis of no breaks versus the alternative hypothesis of presence of at least one break); the test statistics checking for one break against zero -F(1|0)-, two breaks against one -F(2|1)-, and so on until testing five breaks against four -F(5|4). Panel B contains the Bai and Perron statistics for the test run on the regression model selected with the LWZ information criterion. Finally, Panel C reports the test conclusive results, with initial and final days for each regime, the corresponding 90% and 95% confidence intervals and the regime mean. For each Panel, results are provided for both the VIX and standardized VIX time series, listing robust standard errors within brackets; here the significance level is: 1%***, 5%**, 10%*.

Results underpin the presence of four structural breaks, and five regimes at the mean level in total. We were able to link those structural breaks with four major events that can explain the change in the regime that is, in chronological order: the 2008 global financial crisis; the 2010 European Sovereign debt crisis triggered by the Greek situation; the 2012 end of uncertainty in the European markets due to new expansionary monetary policies; the 2016 Brexit and related uncertainty period. We were therefore able to highlight five regimes within the available sample of 2791 observations: Regime 1 starting in Jan:2007 and ending in Sept:2008 (419 obs.); Regime 2 starting in Sept:2008 and ending in Jul:2010 (463 obs.); Regime 3 starting in Jul:2010 and ending in March:2012 (422 obs.); Regime 4 starting in March:2012 and ending in June:2016 (1072 obs.); Regime 5 starting in June:2016 and ending in Feb:2017 (414 obs.). Regime 1 and 4 can be viewed as periods of "regular" volatility, Regime 2 and 3 are high volatility periods while Regime 5 is the sole low volatility period. Figure 1 provides a visual illustration: the VIX time series (in blue) was divided into five bands whose amplitude relates to the regimes length and bound corresponding to the events determining the breaks.

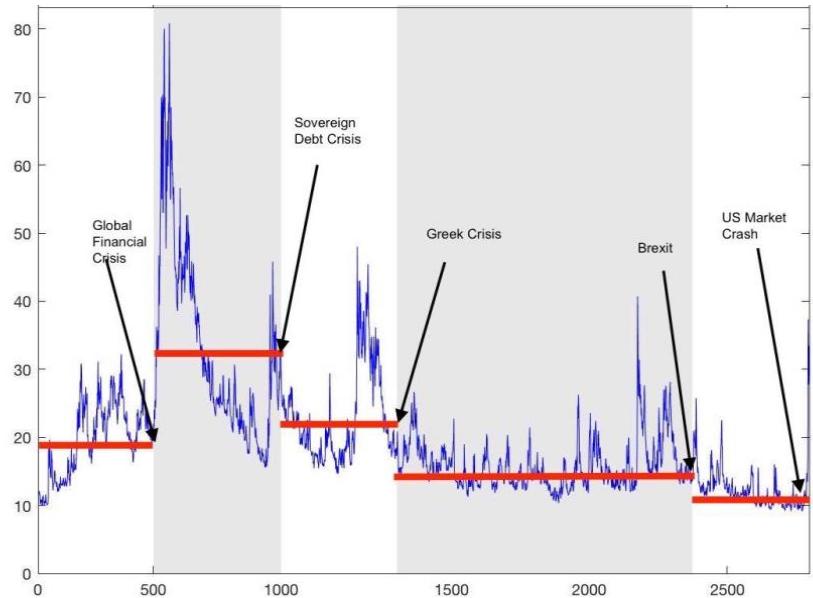

**Figure 1**: The VIX time series with highlighted structural breaks: values are percentage points standing for the expected range of movement in the S&P 500 index over the next year, at a 68% confidence level (i.e. one standard deviation of the normal probability curve). On the x-axis integer values correspond to days in the range: 3 Jan 2007 (t=0)- 1 February 2018 (t=2791).

During the period of interest for this work (Jan 2007 – Feb 2018), the VIX behavior seemed to have been pulled at the extremes more than as previously observed: this is mostly due to the presence in our sample of the 2008 financial crisis; however, also including one of the highest high (80.86, 20 Nov 2008) and one of the lowest low (9.14, 3 Nov 2017) of VIX whole history played an important role. Following (Whaley, 2000), we try to assess this evidence in a more quantitative way by looking at median and percentile ranges for daily closing VIX levels, in Table 2.

**Table 2**: Median and percentile ranges for daily closing VIX levels.

| Year | Obs. | Percentile | | | | | | | Normal Ranges | |
| --- | --- | --- | --- | --- | --- | --- | --- | --- | --- | --- |
| | | 5% | 10% | 25% | 50% | 75% | 90% | 95% | 50% | 90% |
| Full | 2797 | 10.614 | 11.680 | 13.488 | 17.010 | 22.963 | 30.664 | 39.934 | 9.475 | 29.320 |
| Regime 1 | 419 | 10.754 | 12.288 | 14.935 | 20.410 | 23.818 | 26.348 | 27.780 | 8.883 | 17.027 |
| Regime 2 | 463 | 17.522 | 18.808 | 23.065 | 28.580 | 41.938 | 53.680 | 62.928 | 18.873 | 45.406 |
| Regime 3 | 422 | 15.950 | 16.511 | 17.950 | 20.800 | 25.710 | 33.128 | 36.946 | 7.760 | 20.996 |
| Regime 4 | 1072 | 12.071 | 12.477 | 13.400 | 14.850 | 17.085 | 20.536 | 23.078 | 3.685 | 11.007 |
| Regime 5 | 414 | 9.598 | 9.870 | 10.458 | 11.660 | 13.190 | 15.580 | 18.398 | 2.733 | 8.800 |

Looking at Table 2, for the whole sample (first row, with the label "Full") and for each regime under examination, we report the number of observations, seven percentiles in ascendant order from 5% to 95%, and two Normal ranges: the 50% (90%) one is obtained by subtracting the 25% (95%) percentile from the 75% (5%) percentile. The 50% percentile is the median VIX closing value: for the full sample it is at 17.01%, i.e. the VIX, with a probability of 0.5, for the next year should maintain between 13.48% and 22.96%. However, as highlighted in (Whaley, 2009) these results are meaningless if not compared with historical values. To such aim, we made two comparisons. First, we compared our findings with those in (Whaley, 2000), who examined the VIX in the period 1986-1999. Whaley's sample includes a financial crisis (1987) and a market mini-crash (1989), so we expect to find in our samples quite similar median and percentile ranges. To support this assertion, we may observe that in the period studied by Whaley the VIX has a median value of 17.30%, with the probability of 50% of lying in the range 12.97% - 23.04%; these values are very close to those examined in our case. Nevertheless, when moving to Normal ranges (in the latest right-hand columns of Table 2), the two data samples show distinctive features. In particular, the 2008 financial crisis is clearly more volatile than the 1987 turmoil analyzed by Whaley: for the 1987 event, the 50% and 90% Normal ranges were 6.13% and 38.45%, respectively; during the 2008 event, on the other hand, the corresponding Normal ranges are: 18.87% and 45.40%. Second, we compared our results with those in (Sarwar, 2012b) who focused on the relations between the VIX and BRIC markets indexes, but in a different period. Sarwar's sample goes from 1993 to 2007, and includes the markets fall due to 9/11 facts. However, the features diversity among the samples is stated in the number of structural breaks: two in the Sarwar's case against a break every 2 year on average, in our sample.

## 2.3 Descriptive Statistics

We are now going to provide descriptive statistics for the time series analyzed in the paper: relevant values are highlighted in Table 3 for VIX, the changes in VIX (cVIX thereinafter) as well as US and BRIC stock index returns; statistics are reported for both the whole observation period and for all the identified regimes. The cVIX values are scaled by a factor of 100.

**Table 3**: Descriptive Statistics for VIX, cVIX, the US and BRIC stock market indexes. For the Augmented Dickey-Fuller (ADF) test, test statistics with significance are reported.

| Variable | Mean | StdDev | Min | Max | $\rho(1)$ | $\rho(2)$ | $\rho(3)$ | ADF test |
|---|---|---|---|---|---|---|---|---|
| **Full Period – 2007-2018** | | | | | | | | |
| VIX | 19.7806 | 9.6665 | 9.1400 | 80.8600 | 0.9807 | 0.9663 | 0.9546 | -2.2586* |
| cVIX | 0.0000 | 0.0189 | -0.1736 | 0.1654 | 0.9807 | 0.9663 | 0.9546 | -60.0143*** |
| SPX | 0.0002 | 0.0126 | -0.0947 | 0.1096 | -0.1045 | -0.0556 | 0.0360 | -58.5676*** |
| IBOV | 0.0002 | 0.0172 | -0.1412 | 0.1368 | -0.0067 | -0.0213 | -0.0363 | -53.1141*** |
| IMOEX | 0.0001 | 0.0203 | -0.2066 | 0.2523 | -0.0018 | 0.0098 | -0.0324 | -52.8806*** |
| BSESN | 0.0003 | 0.0139 | -0.1160 | 0.1599 | 0.0901 | -0.0278 | -0.0210 | -48.1870*** |
| SHSEC | 0.0001 | 0.0167 | -0.0967 | 0.0903 | 0.0021 | 0.0059 | 0.0423 | -52.6678*** |
| **Regime 1 – 2007-2008** | | | | | | | | |
| VIX | 19.7580 | 5.3622 | 9.8900 | 32.2400 | 0.9548 | 0.9294 | 0.9074 | -0.5211 |
| cVIX | 0.0003 | 0.0157 | -0.0699 | 0.0716 | 0.9548 | 0.9294 | 0.9074 | -25.9467*** |
| SPX | -0.0003 | 0.0116 | -0.0353 | 0.0415 | -0.1609 | -0.0015 | -0.0345 | -23.9978*** |
| IBOV | 0.0005 | 0.0175 | -0.0686 | 0.0614 | -0.0321 | 0.0294 | -0.0563 | -21.0783*** |
| IMOEX | -0.0007 | 0.0183 | -0.1296 | 0.0564 | -0.0969 | -0.0090 | 0.0772 | -22.4771*** |
| BSESN | 0.0001 | 0.0187 | -0.0776 | 0.0641 | 0.0721 | 0.0319 | -0.0516 | -18.9525*** |
| SHSEC | -0.0004 | 0.0246 | -0.0967 | 0.0889 | -0.0522 | -0.0037 | 0.1304 | -21.4919*** |
| **Regime 2 – 2008-2010** | | | | | | | | |
| VIX | 32.9862 | 13.7895 | 15.5800 | 80.8600 | 0.9724 | 0.9518 | 0.9394 | -0.8747 |
| cVIX | 0.0000 | 0.0320 | -0.1736 | 0.1654 | 0.9724 | 0.9518 | 0.9394 | -24.4204*** |
| SPX | -0.0003 | 0.0225 | -0.0947 | 0.1096 | -0.1255 | -0.1262 | 0.1186 | -24.3503*** |
| IBOV | 0.0005 | 0.0265 | -0.1412 | 0.1368 | -0.0037 | -0.1066 | -0.0418 | -21.5482*** |
| IMOEX | 0.0001 | 0.0389 | -0.2066 | 0.2523 | 0.0192 | 0.0242 | -0.0613 | -21.0827*** |
| BSESN | 0.0004 | 0.0220 | -0.1160 | 0.1599 | 0.0992 | -0.0807 | -0.0106 | -19.4606*** |
| SHSEC | 0.0001 | 0.0198 | -0.0698 | 0.0903 | 0.0048 | 0.0235 | 0.0294 | -21.4322*** |
| **Regime 3 – 2010-2012** | | | | | | | | |
| VIX | 22.9438 | 6.7439 | 14.6200 | 48.0000 | 0.9512 | 0.9210 | 0.8846 | -1.1723 |
| cVIX | -0.0002 | 0.0207 | -0.1294 | 0.1600 | 0.9512 | 0.9210 | 0.8846 | -25.1426*** |
| SPX | 0.0006 | 0.0126 | -0.0690 | 0.0463 | -0.1083 | 0.0990 | -0.1492 | -22.7963*** |
| IBOV | 0.0001 | 0.0138 | -0.0843 | 0.0498 | -0.0407 | 0.0653 | -0.0542 | -21.3226*** |
| IMOEX | 0.0004 | 0.0147 | -0.0814 | 0.0407 | 0.0612 | -0.0187 | -0.0212 | -19.2619*** |
| BSESN | 0.0000 | 0.0117 | -0.0421 | 0.0352 | 0.1015 | 0.0156 | -0.0023 | -18.5087*** |
| SHSEC | 0.0000 | 0.0120 | -0.0529 | 0.0313 | -0.0290 | 0.0514 | -0.0280 | -21.0790*** |
| **Regime 4 – 2012-2016** | | | | | | | | |
| VIX | 15.8058 | 3.5255 | 10.3200 | 40.7400 | 0.9276 | 0.8565 | 0.7932 | -1.4506 |
| cVIX | 0.0000 | 0.0133 | -0.0570 | 0.1271 | 0.9276 | 0.8565 | 0.7932 | -33.0528*** |
| SPX | 0.0004 | 0.0083 | -0.0402 | 0.0383 | -0.0038 | -0.0177 | -0.0065 | -32.7749*** |
| IBOV | -0.0002 | 0.0146 | -0.0499 | 0.0639 | 0.0284 | 0.0041 | -0.0154 | -31.7920*** |
| IMOEX | 0.0002 | 0.0133 | -0.1142 | 0.1136 | -0.0668 | -0.0055 | -0.0282 | -34.9605*** |
| BSESN | 0.0004 | 0.0095 | -0.0612 | 0.0337 | 0.0875 | -0.0147 | -0.0217 | -29.9089*** |
| SHSEC | 0.0002 | 0.0159 | -0.0887 | 0.0604 | 0.0562 | -0.0034 | -0.0187 | -30.9209*** |
| **Regime 5 – 2016-2018** | | | | | | | | |
| VIX | 12.2059 | 2.4864 | 9.1400 | 25.7600 | 0.9033 | 0.8245 | 0.7516 | -0.9028 |
| cVIX | 0.0000 | 0.0109 | -0.0510 | 0.0851 | 0.9033 | 0.8245 | 0.7516 | -22.2678*** |
| SPX | 0.0007 | 0.0054 | -0.0366 | 0.0220 | -0.0684 | -0.0346 | -0.0693 | -21.2633*** |
| IBOV | 0.0012 | 0.0126 | -0.0921 | 0.0391 | -0.0641 | 0.0924 | -0.0670 | -21.4489*** |
| IMOEX | 0.0004 | 0.0080 | -0.0246 | 0.0281 | 0.1883 | -0.0333 | 0.0042 | -16.7639*** |
| BSESN | 0.0007 | 0.0063 | -0.0257 | 0.0187 | 0.0872 | -0.0534 | 0.0534 | -18.3985*** |
| SHSEC | 0.0004 | 0.0062 | -0.0250 | 0.0249 | 0.0209 | -0.0750 | 0.0430 | -19.7341*** |

For each index we show: Mean, Standard Deviation (StDev), Minimum (Min), Maximum (Max), autocorrelations at various lags ($\rho(1)$-$\rho(3)$) and the test statistics for the Augmented Dickey-Fuller (ADF) test. Results foster the splitting identified by the Bai and Perron test in Sec. 2.2. The volatility, in fact, is time-changing: the VIX mean value is 19.75% during Regime 1 but moves to 32.98% (due to the 2008 financial crisis) during Regime 2 and goes back to the original Regime 1 level in the following years; finally, during the last period (Regime 5) it settles to its lowest level. The dualism between Regime 2 and Regime 5 is also supported by considering that while in the former the VIX reached its highest peak (80.86%), in the latter, the VIX reverted to its lowest value ever (9.14%). Finally, most of the series along the full sample and the sub-samples show a slightly negative autocorrelation. This is in line with the findings of (Fleming et al., 1995): the VIX does not exhibit seasonality patterns, while changes in VIX are slightly autocorrelated.

The remarks in previous rows support the decision of using cVIX instead of VIX to estimate the intertemporal relationships between the VIX and stock market indexes. First, intuitively, changes into expected volatility reflect what investors may be worried from. Second, a regression involving raw VIX values and prices would lead to spurious estimations: from last column in Table 3, in fact, we can see that while cVIX and index returns are stationary, both the VIX and index prices are not.

# 3 Methodology

Inspired by the works of (Fleming et al., 1995) first and (Sarwar, 2012a,b) after, we applied the (Fleming et al., 1995) multivariate regression model to assess the intertemporal relationships among the changes in VIX and the stock market returns indexes of US and BRIC countries.

The model has a long and well-established track in investigating the relationship between market indexes and the VIX. Table 4 lists all the works using this model to address the VIX fear gauge role. For each work, we showed the dependent and independent variables, the countries of interest, the period of analysis and the estimation methodology.

**Table 4**: Applications of the (Fleming et al., 1995)'s multivariate regression model to assess the fear gauge role of the VIX index.

| Reference | Dependent Var. | Independent Var. | Countries | Period | Est. Method. |
|---|---|---|---|---|---|
| (Fleming et al., 1995) | cVIX | Stock Index Returns | US | 1986-1992 | GMM |
| (Sarwar, 2012a) | cVIX | Stock Index Returns | US | 1992-2011 | GMM |
| (Sarwar, 2012b) | cVIX | Stock Index Returns | US and BRIC | 1993-2007 | GMM |
| (Sarwar, 2014) | cVIX | Stock Index Returns | Europe | 1998-2013 | GMM |
| (Sarwar, 2017) | cVIX | Changes in Volatility of T-note and precious metals | US | 2004-2014 | GMM |
| (Sarwar and Khan, 2017) | Stock Index Returns | cVIX | Latin America | 2003-2014 | GMM |

With the exclusion of (Sarwar and Khan, 2017), the remaining papers share the same aim, as they investigate whether the VIX index is a fear gauge for several stock market indexes during different periods. Conversely, (Sarwar and Khan, 2017) examine the effects of VIX on Latin America stock market indexes, by swapping the VIX from dependent to independent variable. Finally, with respect to the estimation procedure, the model always relies on the Generalized Method of Moments (GMM); this is justified by at least two motivations: (*i*) the autocorrelation in stock market index returns; (*ii*) the heteroscedasticity and autocorrelation in the VIX changes, as noted in (Fleming et al., 1995). Indeed, (Fleming et al., 1995) is the only case where the GMM is made more robust by considering Parzen weights (Gallant, 1987) for the covariance matrix estimation and Andrew's method of automatic bandwidth selection (Andrews, 1991).

Going back to the model, it allows considering five temporal relationships at once: let us denote by *t* the reference trading time, then *t-1* and *t-2* identify two lags, while *t+1* and *t+2* indicate two leads. Index returns for each examined market were then regressed for any of the above lag and lead against current changes in the VIX value. For the generic country *s*, we therefore have:

$$cVIX_t = \alpha + \sum_{i=-2}^{2} \beta_{s,t+i} R_{s,t+i} + \beta_{|s|,t} |R_{s,t}| + \varepsilon_t. \qquad (1)$$

Here $cVIX_t$ is the change in the VIX at time *t*; $\alpha$ is the intercept of the regression, $R_{s,t+i}$ is the index return for the market *s* at time $t+i$ (*i*=-2,-1,0, 1,2); $|R_{s,t}|$ is the current absolute return and $\varepsilon_t$ is the error term. Lags and leads in the model help managing the issue of time shift, as VIX and BRIC market indexes are traded at different times.

The richness of existing ties between changes in the VIX and the market indexes is fully contained in (1). The model, in fact, captures the simultaneous relationship between cVIX$_t$ and the s-th market index return through the parameter $\beta_{s,t}$, while delayed values, ruled out by the $\beta_{s,t+i}$ (*i*≠0), explain both ex-ante and ex-post relationships among cVIX$_t$ and stock market returns. In general, the fear gauge role of the VIX index is assessed when $\beta_{s,t+i}$ (*i*≠0) assume negative values: we expect to find a kind of inverse relationship: a decrease (rise) in the market index value should be accompanied by an increase (decrease) in market expected volatility (Fleming et al., 1995).

The importance of the parameters depends on the number of trading hours shared between the VIX index and the country stock index. In fact, when they are fully matching (as in the case of S&P500 index), $\beta_{s,t}$ captures almost the whole relationship, while lead and lag parameters are close to zero (Fleming et al., 1995, Sarwar, 2012a,b). On the contrary, in case of low or even in the absence of overlapping trading hours, when the country index closes before (after) VIX opening, the delayed values assume greater importance - and a higher value in absolute terms - since they capture a kind of backward/forward shift in the VIX effect. Nevertheless, (Sarwar, 2012b) find that the relationship between BRIC stock markets index which trade before the VIX closing time and cVIX$_t$ is well captured by $\beta_{s,t}$ and $\beta_{s,t+1}$.

To conclude, we examine the role of $\beta_{|s|,t}$, that is a proxy of market sentiment: when negative (positive), the market volatility should equivalently tend to decrease (increase). Besides, for *i*=0, we use $\beta_{|s|,t}$ also to calculate the sums $\beta_s^+ = \beta_{s,t} + \beta_{|s|,t}$ and $\beta_s^- = \beta_{s,t} - \beta_{|s|,t}$, which are the contemporaneous responses to positive and negative market shifts and give indications about the eventual asymmetry between cVIX$_t$ and the stock market returns. In order to evaluate which of them prevail on the other, we will compare the corresponding absolute values.

As final note, we underline that in (1) we only inserted $\beta_{|s|,t}$ without considering any lag or lead. This choice obeys basically to two motivations: first, (Fleming et al., 1995; Sarwar, 2012a,b) found lag and lead absolute parameters not-significant; second, we want to avoid the over-parameterization of the model.

## 4 Results

In order to verify whether the VIX index has been a fear gauge for the US and the BRIC stock market indexes during the period Jan 2007 – Feb 2018, we start at first by looking at the correlation between the cVIX and the intertemporal stock index returns values; we will then move to estimate the multivariate regression model.

Table 5 lists the correlation between cVIX$_t$ and values of indexes returns at various lags and leads $t+i$, ($i$=-2,-1,0,1,2) for the whole sample (Column 2) as well as for the regimes identified in Sec. 2 (Columns 3 to 7).

**Table 5**: Correlation between cVIX$_t$ and indexes returns at time $t+i$, $i$ = -2,-1,0,1,2.

| Correlation Pair | Full | 2007-2008 | 2008-2010 | 2010-2012 | 2012-2016 | 2016-2018 |
|---|---|---|---|---|---|---|
| SPX(-2)-cVIX(0) | 0.0594*** | 0.0281 | 0.1112** | -0.0720 | 0.0437 | 0.1159** |
| SPX(-1)-cVIX(0) | 0.1091*** | 0.2015*** | 0.1046** | 0.1409*** | 0.0483 | 0.0762 |
| SPX(0)-cVIX(0) | -0.8346*** | -0.8558*** | -0.8421*** | -0.8675*** | -0.8394*** | -0.7762*** |
| SPX(+1)-cVIX(0) | 0.0949*** | 0.1621*** | 0.1238*** | 0.1508*** | -0.0663** | 0.0579 |

| | | | | | |
|---|---|---|---|---|---|
| SPX(+2)-cVIX(0) | 0.0684*** | -0.0040 | 0.1656*** | -0.0973** | 0.0081 | 0.0516 |
| IBOV(-2)-cVIX(0) | 0.0261*** | -0.0610 | 0.0734 | -0.1120** | 0.0603** | 0.0433 |
| IBOV(-1)-cVIX(0) | 0.0765*** | 0.2076*** | 0.0852* | 0.1393*** | -0.0359 | 0.0572 |
| IBOV(0)-cVIX(0) | -0.5814*** | -0.6194*** | -0.7319*** | -0.6640*** | -0.3643*** | -0.3453*** |
| IBOV(+1)-cVIX(0) | -0.0143*** | 0.0092 | -0.0176 | 0.0479 | -0.0612** | -0.0063 |
| IBOV(+2)-cVIX(0) | 0.0438*** | -0.0182 | 0.1098** | -0.0295 | 0.0029 | 0.0416 |
| IMOEX(-2)-cVIX(0) | 0.0019*** | -0.0009 | -0.0026 | 0.0107 | -0.0167 | 0.1196** |
| IMOEX(-1)-cVIX(0) | 0.0328*** | 0.0745 | 0.0250 | 0.0697 | 0.0238 | -0.0101 |
| IMOEX(0)-cVIX(0) | -0.3022*** | -0.1658*** | -0.3422*** | -0.4578*** | -0.2466*** | -0.1529*** |
| IMOEX(+1)-cVIX(0) | -0.1777*** | -0.2307*** | -0.2266*** | -0.0298 | -0.1130*** | -0.2223*** |
| IMOEX(+2)-cVIX(0) | -0.0207*** | 0.0283 | -0.0226 | -0.1227** | 0.0122 | 0.0178 |
| BSESN(-2)-cVIX(0) | 0.0237*** | -0.1023** | 0.0657 | -0.0048 | 0.0560* | 0.0472 |
| BSESN(-1)-cVIX(0) | 0.0570*** | 0.1533*** | 0.0429 | 0.0470 | 0.0548* | -0.0484 |
| BSESN(0)-cVIX(0) | -0.2823*** | -0.0540 | -0.4358*** | -0.1602*** | -0.2944*** | -0.1795*** |
| BSESN(+1)-cVIX(0) | -0.2287*** | -0.3749*** | -0.1929*** | -0.2326*** | -0.2169*** | -0.2552*** |
| BSESN(+2)-cVIX(0) | -0.0182*** | -0.0372 | 0.0176 | -0.0623 | -0.0661** | 0.0073 |
| SHSEC(-2)-cVIX(0) | 0.0031*** | -0.0583 | 0.0111 | -0.0123 | 0.0522* | -0.0371 |
| SHSEC(-1)-cVIX(0) | 0.0279*** | 0.0308 | 0.0606 | 0.0051 | 0.0108 | -0.0131 |
| SHSEC(0)-cVIX(0) | -0.0838*** | -0.0039 | -0.1193** | -0.1055** | -0.1362*** | 0.0514 |
| SHSEC(+1)-cVIX(0) | -0.1525*** | -0.1554*** | -0.1576*** | -0.1785*** | -0.1757*** | -0.1935*** |
| SHSEC(+2)-cVIX(0) | -0.0099*** | -0.0337 | 0.0741 | -0.0679 | -0.0649** | -0.0805* |

During the full period (Column 2), the correlation between current index values and VIX was negative for all the indexes and always statistical significant. In particular, the SPX has the lower value (-0.83); followed in ascendant order by IBOV (-0.58); IMOEX (-0.30); BSESN (-0.28) and SHSEC (-0.08). This result is coherent with the shift in trading hours highlighted in Section 2: the higher the overlap is, the more contemporaneous values are negatively correlated. Moreover, in the whole sample also lead and lag values are always significant for all the indexes under consideration. In detail, for the SPX the values are negative and sensitively small, thus supporting the findings in (Sarwar, 2012b); for BRIC indexes lag values are small and positive, and hence denote that past stock returns are not influencing the VIX. On the other hand,

20lead values are mostly negative and small: the majority of the effect on VIX should be therefore explained by indexes contemporaneous values, as observed in (Sarwar, 2012b). In detail, IBOV, whose trading hours are nearly overlapping to those of VIX, shows positive and small value at lead 2 IBOV(+2), as the SPX. The remaining three indexes, on the contrary, have negative but small correlation values at lead *2*, and negative and higher correlation at lead *1*, whose magnitude is similar to the contemporaneous correlation value. We can therefore state that the VIX effect is therefore captured by the contemporaneous correlation values, as well as at lead *1* and, weekly at lead *2*.

Moving to sub-periods, the contemporaneous relation changes from the full period case: for the SPX, the correlation has the same magnitude across all the sub-samples with the exception of the last one (2016-2018), characterized by a lower level of volatility. Again, for SPX, the correlation is always statistically significant at lag 1 as well as at lead 1: SPX(+1) and SPX(-1) are positive and small, especially during less volatile periods. The remaining lags and leads offer mixed result. We can therefore state that during sub-periods, the statistical significance of correlation helps at identifying where the relationships lies, i.e. in the current values as well as at lead and lag 1. Regarding BRIC indexes, trading time plays a crucial role in determining when the correlation is stronger. In detail, the IBOV, which is the index with the highest overlapping in trading hours with VIX, shows most relevant correlation value at time *t*, with small values elsewhere. BSESN, IMOEX and SHSEC, on the other hand, show most significant correlation at lead *t+1*; nevertheless, correlation at time *t* is still negative. In conclusion, the correlation between BRIC indexes and cVIX (with the IBOV as exception) lies mainly in the index values at both time $t$ and $t+1$, with significant and generally negative values, while lagged correlations are small and mostly not statistically significant. Correlation is stronger during the 2008 financial crisis, while it is softened during more recent periods, and especially in the latest one.

We are now ready to apply the multivariate regression model: Table 6 shows the intertemporal relationships for the US and BRIC stock markets estimated with (1) through the GMM procedure. The columns report the values for estimated parameters: $\hat{\beta}_{s,t+i}$ (with $i$=-2,-1,0,1,2), $\hat{\beta}_{|s|,t}$, $\hat{\beta}_s^-$ and $\hat{\beta}_s^+$, whose meaning has been already explained in Sec. 3. Standard errors are within rounded brackets and all significant at 99% confidence level, therefore we omitted to mark them by the conventionally used *** notation.

**Table 6**: Intertemporal relationships for the US and BRIC stock markets.

| Period | Intercept | $\hat{\beta}_{s,t-2}$ | $\hat{\beta}_{s,t-1}$ | $\hat{\beta}_{s,t}$ | $\hat{\beta}_{s,t+1}$ | $\hat{\beta}_{s,t+2}$ | $\hat{\beta}_{|s|,t}$ | $\hat{\beta}_s^+$ | $\hat{\beta}_s^-$ |
|---|---|---|---|---|---|---|---|---|---|
| **Panel A: S&P500** | | | | | | | | | |
| FULL | -0.0007 | 0.0324 | 0.0571 | -1.2406 | 0.0115 | 0.0332 | 0.1231 | 1.1175 | 1.3637 |
|  | (-1.9570) | (0.9109) | (1.8867) | (-22.5757) | (0.3256) | (0.8744) | (2.6079) | | |
| 07-08 | -0.0008 | 0.0636 | 0.1011 | -1.1261 | 0.0370 | -0.0147 | 0.0969 | 1.0291 | 1.2230 |
|  | (-1.3552) | (1.4375) | (2.5713) | (-17.9311) | (1.1666) | (-0.3791) | (1.2543) | | |
| 08-10 | -0.0019 | 0.0103 | 0.0095 | -1.1700 | 0.0367 | 0.0966 | 0.1097 | 1.0603 | 1.2796 |
|  | (-1.7129) | (0.1956) | (0.2032) | (-13.7895) | (0.6863) | (1.7589) | (1.5140) | | |
| 10-12 | -0.0021 | 0.0808 | 0.1098 | -1.3854 | 0.0851 | -0.0054 | 0.2952 | 1.0902 | 1.6806 |
|  | (-2.9990) | (2.4521) | (2.1495) | (-13.6190) | (1.4024) | (-0.1106) | (4.4734) | | |
| 12-16 | -0.0009 | 0.0711 | 0.0915 | -1.3447 | -0.1095 | -0.0074 | 0.2260 | 1.1188 | 1.5707 |
|  | (-1.9828) | (2.1803) | (3.0278) | (-21.6796) | (-1.5105) | (-0.2383) | (2.7295) | | |
| 16-18 | -0.0002 | 0.1320 | 0.2711 | -1.5407 | -0.1439 | -0.0782 | 0.3019 | 1.2388 | 1.8426 |
|  | (-0.3395) | (1.2612) | (1.6259) | (-14.1543) | (-1.7980) | (-1.0247) | (2.0911) | | |
| **Panel B: IBOV** | | | | | | | | | |
| FULL | -0.0014 | 0.0182 | 0.0924 | -0.6399 | -0.0286 | 0.0439 | 0.1259 | 0.5141 | 0.7658 |
|  | (-2.9530) | (0.6685) | (3.8774) | (-10.1428) | (-0.9728) | (1.2797) | (3.1362) | | |
| 07-08 | 0.0000 | -0.0262 | 0.1683 | -0.5450 | -0.0202 | 0.0079 | 0.0379 | 0.5071 | 0.5829 |
|  | (-0.0200) | (-0.6428) | (3.4368) | (-10.8693) | (-0.6342) | (0.2666) | (0.4977) | | |
| 08-10 | -0.0024 | -0.0029 | 0.1164 | -0.8761 | -0.0180 | 0.0545 | 0.1599 | 0.7163 | 1.0360 |
|  | (-1.8589) | (-0.0750) | (3.1690) | (-11.9560) | (-0.3048) | (0.9231) | (2.3158) | | |
| 10-12 | -0.0031 | -0.0788 | 0.1896 | -0.9592 | -0.0017 | 0.0254 | 0.2988 | 0.6605 | 1.2580 |
| | (-3.3611) | (-0.5816) | (1.7443) | (-7.6017) | (-0.0239) | (0.5402) | (3.7937) | | |
| 12-16 | -0.0012 | 0.0607 | -0.0270 | -0.3356 | -0.0496 | 0.0055 | 0.0985 | 0.2371 | 0.4341 |
| | (-2.2653) | (2.2593) | (-0.9669) | (-6.5627) | (-1.5129) | (0.1554) | (1.8625) | | |
| 16-18 | 0.0005 | 0.0391 | 0.0919 | -0.3261 | -0.1043 | 0.0922 | -0.0314 | 0.3575 | 0.2946 |
| | (0.5916) | (0.8788) | (1.2945) | (-4.1770) | (-1.3678) | (1.8984) | (-0.3096) | | |
| **Panel C: IMOEX** | | | | | | | | | |
| FULL | -0.0011 | 0.0028 | 0.0371 | -0.2831 | -0.1663 | -0.0120 | 0.0972 | 0.1859 | 0.3803 |
| | (-2.9838) | (0.1116) | (1.6513) | (-7.5418) | (-4.6826) | (-0.3487) | (3.1047) | | |
| 07-08 | -0.0001 | 0.0197 | 0.0491 | -0.1578 | -0.2159 | -0.0044 | 0.0106 | 0.1471 | 0.1684 |
| | (-0.0727) | (0.5811) | (1.3867) | (-4.6218) | (-4.5297) | (-0.0766) | (0.2160) | | |
| 08-10 | -0.0036 | 0.0041 | 0.0376 | -0.2743 | -0.1794 | 0.0091 | 0.1476 | 0.1268 | 0.4219 |
| | (-2.7618) | (0.0761) | (0.9363) | (-4.7738) | (-3.1390) | (0.1673) | (3.0925) | | |
| 10-12 | -0.0013 | -0.0106 | 0.1464 | -0.6360 | 0.0153 | -0.1858 | 0.1286 | 0.5075 | 0.7646 |
| | (-0.6240) | (-0.1701) | (1.6543) | (-4.8420) | (0.1793) | (-1.5744) | (0.6242) | | |
| 12-16 | -0.0006 | -0.0223 | 0.0018 | -0.2541 | -0.1293 | 0.0022 | 0.0651 | 0.1890 | 0.3193 |
| | (-1.2840) | (-0.9328) | (0.0600) | (-5.7642) | (-3.8487) | (0.0629) | (1.5169) | | |
| 16-18 | 0.0000 | 0.1354 | 0.0366 | -0.2623 | -0.2323 | -0.0026 | 0.0193 | 0.2430 | 0.2815 |
| | (0.0285) | (2.5762) | (0.3455) | (-2.7051) | (-2.5732) | (-0.0421) | (0.1766) | | |
| **Panel D: BSESN** | | | | | | | | | |
| FULL | -0.0014 | 0.0171 | 0.1188 | -0.3749 | -0.2749 | -0.0023 | 0.1669 | 0.2081 | 0.5418 |
| | (-3.4434) | (0.4032) | (2.5724) | (-4.4882) | (-7.3987) | (-0.0692) | (3.8909) | | |
| 07-08 | -0.0001 | -0.1081 | 0.1504 | -0.0294 | -0.3212 | -0.0040 | 0.0327 | 0.0032 | 0.0621 |
| | (-0.1372) | (-2.1601) | (3.0902) | (-0.8341) | (-6.8663) | (-0.0932) | (0.6723) | | |
| 08-10 | -0.0034 | 0.0432 | 0.1148 | -0.6242 | -0.2118 | 0.0116 | 0.2415 | 0.3827 | 0.8657 |
| | (-2.3024) | (0.5660) | (1.0616) | (-4.8832) | (-2.9098) | (0.1405) | (2.7159) | | |
| 10-12 | -0.0033 | 0.0071 | 0.1384 | -0.2601 | -0.3762 | -0.0602 | 0.3319 | 0.0717 | 0.5920 |
| | (-2.2631) | (0.0956) | (1.7779) | (-2.9870) | (-3.6185) | (-0.6583) | (2.0177) | | |
| 12-16 | -0.0019 | 0.0722 | 0.0979 | -0.4012 | -0.2684 | -0.0843 | 0.2927 | 0.1084 | 0.6939 |
| | (-2.2772) | (1.4309) | (2.2187) | (-4.9999) | (-4.2049) | (-1.7562) | (2.3341) | | |
| 16-18 | -0.0005 | 0.0459 | 0.1116 | -0.3309 | -0.3975 | 0.1237 | 0.1674 | 0.1634 | 0.4983 |
| | (-0.4128) | (0.5383) | (1.0538) | (-1.6956) | (-3.1035) | (0.8398) | (0.6954) | | |
| **Panel E: SHSEC** | | | | | | | | | |
| FULL | -0.0010 | 0.0242 | 0.0365 | -0.0904 | -0.1751 | -0.0075 | 0.0973 | 0.0068 | 0.1877 |
| | (-2.6062) | (0.8858) | (1.3466) | (-2.2669) | (-4.9027) | (-0.2586) | (2.7292) | | |
| 07-08 | -0.0005 | -0.0189 | 0.0226 | -0.0001 | -0.0998 | -0.0292 | 0.0368 | 0.0368 | 0.0369 |
| | (-0.4463) | (-0.8191) | (0.7957) | (-0.0021) | (-2.6227) | (-0.8665) | (0.6984) | | |
| 08-10 | -0.0026 | 0.0353 | 0.1049 | -0.1860 | -0.2727 | 0.1247 | 0.1892 | 0.0032 | 0.3752 |
| | (-1.5749) | (0.3467) | (1.2661) | (-1.4883) | (-4.3338) | (1.4386) | (1.8099) | | |
| 10-12 | -0.0043 | 0.0157 | 0.0266 | -0.1539 | -0.3355 | -0.0975 | 0.4594 | 0.3055 | 0.6132 |
| | (-2.6926) | (0.2452) | (0.3782) | (-1.5162) | (-4.4235) | (-1.3286) | (2.4252) | | |
| 12-16 | -0.0007 | 0.0526 | 0.0151 | -0.0975 | -0.1383 | -0.0497 | 0.0682 | 0.0293 | 0.1656 |
| | (-1.3727) | (1.4029) | (0.3973) | (-2.6224) | (-1.9281) | (-0.9616) | (1.3073) | | |
| 16-18 | -0.0006 | 0.0776 | -0.0963 | -0.1409 | -0.2207 | -0.0243 | 0.1743 | 0.0334 | 0.3152 |
| | (-0.7503) | (0.7608) | (-0.8938) | (-1.1495) | (-1.6011) | (-0.2743) | (1.0327) | | |

We first analyze the results for the US market. In all the periods, our findings are aligned to those already discussed in the literature: VIX is a fear gauge because all the estimated contemporaneous parameters are strongly negative. However, putting a magnifying lens on the results related to the 2008 financial crisis gives unexpected outcomes: pre-crisis parameter $(-1.12)$ is higher than during the crisis $(-1.17)$; however, the parameter value after the crisis is surprisingly lower: $(-1.38)$ during the period 2010-12; $(-1.34)$ during 2012-2016 and even lower $(-1.54)$ during 2016-2018, with the lowest value of all, in contrast with correlation results.

For what is concerning the market sentiment $\hat{\beta}_{|SPX|,t}$ in Column 8, the estimated coefficient underlines a positive relationship between the size of a daily stock market move and the contemporaneous daily change in VIX, for all the examined periods, including the full one. This effect is enhanced since 2008 onwards.

To conclude, the last two columns on the right-hand side of Table 6 list the absolute values of $\hat{\beta}^{-}_{S=SPX}$ and $\hat{\beta}^{+}_{S=SPX}$, i.e. the response to negative and positive returns index, respectively. For the full sample, we found negative asymmetry. The asymmetry in response is boosted after the 2008 crisis onwards. Surprisingly, the largest values are during the "low-volatility" period (2016-2018), with a more than doubled response to negative daily index returns.

The relevance of the results is fostered when compared to similar studies run in the past. In fact, starting from the full sample, the contemporaneous value of (-1.25) emphasizes a stronger impact of the VIX over the US stock market: (Fleming et al., 1995) found a value of -0.75 for the period 1986-1992 and (Sarwar, 2012b), who examined the period 1993-2007, found a contemporaneous effect of $(-0.89)$. We can then conclude that the impact of VIX on the SPX has increased in the last decade, especially during and after the 2008 financial crisis. The same conclusion applies to the size parameter and the asymmetric response to negative stock index returns: they are both higher since 2008, and especially during the recent period of "low-volatility" (2016-2018). Moving to BRIC markets, the VIX has a weaker effect on them than as described for the SPX. In fact, looking at the results for the full sample, the IBOV has a contemporaneous coefficient $\hat{\beta}_{IBOV,t}$ almost twice higher than the SPX ($-0.64$ for the IBOV versus $-1.25$ for the SPX); the BSESN values is almost three times higher ($-0.37$); the IMOEX four times higher ($-0.28$); and the SHSEC almost twelve times higher ($-0.08$). The most relevant ties with the VIX are at lead 1, with the only exception of IBOV, but always smaller than those observed for SPX. We can therefore reasonably argue that among the BRIC stock indexes the IBOV is the most influenced by the VIX at contemporaneous time, while the others are ex post conditioned at lead 1 in an equivalent way. Although these values are not as high as for the SPX, however, they are all negative and significant hence highlight the VIX fear gauge role played for all the BRIC stock indexes. Moreover, the hypothesis that the fear gauge effect is maintained through next period returns (as can be seen by looking at lead +1 parameters) is supported for Russia, China and India: their $\hat{\beta}_{s,t+1}$ coefficients are generally equal or higher in absolute terms than corresponding $\hat{\beta}_{s,t}$, so that they carry on cross-market effects. Besides, $\hat{\beta}_{|s|,t}$ is positive; this leads to an asymmetric response of VIX to BRIC stock market indexes although estimated values are not as high as in the SPX case).

Looking at the effects on the sub-samples, results are controversial. In fact, the observation drawn for the SPX about the impact of 2008 financial crisis are now valid only for IMOEX, with the strongest relationship in the more recent period (2016-2018). On the other hand, for the IBOV and the SHSEC the strongest ties with the VIX are during 2008-2010. In the case of BSESN, the relationship peaked straight after the crisis (2010-2012). All the cited indexes show a mean-reverting behavior after 2010 which has been broken after 2016. Indeed, for sub-samples we have an asymmetrically negative response of VIX to BRIC stock market indexes.

Comparing our results with (Sarwar, 2012b), we can therefore confirm the fear gauge role for all BRIC markets, including Russia. Indeed, in the case of Russia we cannot make a direct comparison with the findings of Sarwar using a different stock market index. Moreover, our results highlight that the VIX effect is more pronounced: during the period 1993-2007, Sarwar found the following coefficients for the contemporaneous and the lead $t+1$ parameters, respectively: (-0.02) and (0.007), for SHSEC; (-0.009) and (-0.009) for BSESN; (-0.030) and (0.009) for IBOV. These outcomes are perfectly in line with our findings which underlined an enhancement of ties during and after the 2008 financial crisis.

In conclusion, the VIX is a strong fear gauge for the US market not only during crisis but even after those critical events: this relationship maintains strong also in low volatile timeframes, as demonstrated by analyzing the 2016-2018 period. This strengthen the fear gauge role of the VIX, which is no more bounded to high volatility periods but rather extended towards low-volatility phases. At the same time, the VIX revealed being a strong fear gauge also for the BRIC stock market indexes. This effect is strictly linked with the overlapping amplitude in trading hours: in the case of IBOV, where the overlap is higher, the fear effect is stronger, and it is more pronounced for the contemporaneous coefficient. For China, India and Russia, the effect is evident both in the contemporaneous and in the lead $t+1$ coefficients. With the exception of IBOV, which behaves as the SPX, the BRIC indexes show a peak in their relationship with VIX during the 2008 financial crisis to decrease thereafter. In addition, for all the markets under consideration, we found strong presence of asymmetric responses to negative market returns.

## 5 Conclusion

We investigated the relationships between the CBOE VIX volatility index and the US and BRIC market indexes, with the aim of verifying to which extent the VIX can be still considered a fear gauge for them. In detail, following the work of (Sarwar, 2012b), we picked up the analysis from the point he left off, focusing on the time window from Jan 2007 to Feb 2018, to capture the relations before, during and after the 2008 financial crisis.

Starting from a rigorous statistical analysis of the VIX and changes in VIX time series, we were able to identify four structural breaks in the VIX and consequently to split the whole sample into five regimes. This enhanced the capability of the multivariate regression model suggested in (Fleming et al., 1995) in estimating the relationships between VIX and the US and BRIC stock markets. In fact, the statistical analysis highlighted how during the period Jan 2007 – Feb 2018 the VIX has reached extreme values, thus indicating the greater uncertainty among investors. This uncertainty was captured by the regression model: the VIX has never been such a strong fear gauge for US market than during and after the 2008 financial crisis. Empirical evidence confirms the asymmetry response to negative returns and shows that the role of VIX for the US is tougher than as highlighted in previous works such as (Fleming et al., 1995) and (Sarwar, 2012b) and that even in low volatile periods, like during the biennium 2016 –2018, the VIX promptly reacted to market drawdowns.

Regarding to BRIC markets, our findings are aligned to those in (Sarwar, 2012b), as we are able to state that the VIX is a fear gauge especially for the Brazilian IBOV, and in a more limited way for the Chinese SHSEC and the Indian BSESN. Contradicting Sarwar's findings, VIX is also a fear gauge for the Russian IMOEX. Overlapping in trading hours played a decisive role in the fear transmission: this explains why IBOV (which has six trading hours overlapping with VIX) has in average a 30% stronger reaction to VIX changes than BSESN, SHSEC and IMOEX. Moreover, VIX fear transmission mechanism worked towards BRIC markets mostly during the 2008 financial crisis 2008-2010 and in the post-crisis period (2010-2012). After 2012, the relationships weakened to revert to pre-crisis values. This applies for all BRIC markets with the exception of IBOV, where we underlined a strengthening of the relation with VIX even after 2008, in line with the US market. The asymmetry in response to negative returns is confirmed for all the BRIC countries.

In conclusion, the 2008 financial crisis exacerbated the VIX fear gauge role for the US and also for the BRIC markets. These effects remain strong for the US even in the latest years where the VIX reached the lowest values in its history. On the other hand, the relationships went back to pre-crisis level for the BRIC indexes until the 2016-2018 period. In addition, our work also supports findings in previous about the relationships between VIX and BRIC markets, thus consecrating the role of VIX as fear gauge for Brazil, China and India and also Russia.